\documentclass[aps,twocolumn,groupedaddress,showpacs,amssymb,prl,floatfix,preprintnumbers]{revtex4}
\usepackage{graphicx,color}

\newcommand{\ybcox}      {${\rm Y} {\rm Ba}_{2} {\rm Cu}_{3} {\rm O}_{6+y}$}

\newcommand{\as}         {$^{75}$As}

\newcommand{\etal}      {{\it et~al}.}
\newcommand{\LaOFFeAs}       {${\rm La} {\rm O}_{0.9} {\rm F}_{0.1} {\rm Fe} {\rm As}$}

\newcommand{\LaOFFeAsx}       {${\rm La} {\rm O}_{1-x} {\rm F}_{x} {\rm Fe} {\rm As}$}

\newcommand{\slrrt}     {$(T_1T)^{-1}$}
\newcommand{\slrr}      {$T_1^{-1}$}

\newcommand{\slrrtext}  {spin lattice relaxation rate}

\definecolor{orange}{rgb}{0.7,0.6,0.1}
\definecolor{violett}{rgb}{0.5,0.0,0.5}

\begin{document}

\thispagestyle{myheadings}

\title{\as\ NMR studies of superconducting \LaOFFeAs}

\author{H.-J. Grafe$^{1}$, D. Paar$^{1,2}$, G. Lang$^{1}$, N. J. Curro$^{3}$, G. Behr$^1$, J. Werner$^1$,
 J. Hamann-Borrero$^1$, C. Hess$^1$, N. Leps$^1$, R. Klingeler$^1$, B. B\"uchner$^1$}
 \affiliation{$^1$IFW Dresden, Institute for Solid State Research, P.O. Box 270116, D-01171 Dresden, Germany\\
 $^2$Department of Physics, Faculty of Science, University of Zagreb, P. O. Box 331, HR-10002 Zagreb,
 Croatia\\
 $^3$Department of Physics, University of California, Davis, CA 95616, USA}

\date{\today}

\begin{abstract}
We have performed \as\ Nuclear Magnetic Resonance (NMR)
measurements on aligned powders of the new \LaOFFeAs\
superconductor. In the normal state, we find a strong temperature
dependence of the spin shift and Korringa behavior of the
\slrrtext. In the superconducting state, we find evidence for line
nodes in the superconducting gap and spin-singlet pairing. Our
measurements reveal a strong anisotropy of the spin lattice
relaxation rate, which suggest that superconducting vortices
contribute to the relaxation rate when the field is parallel to
the $c$-axis but not for the perpendicular direction.
\end{abstract}


\maketitle

The discovery of superconductivity in the iron-oxypnictide
compounds has been focus of tremendous interest in recent months
\cite{LaOFFeAsdiscovery,LaOFFeAsNature}. These compounds have
transition temperatures up to 55 K, yet the conventional BCS
picture of a phonon-mediated pairing mechanism appears to be
untenable due to the weak electron-phonon coupling
\cite{Boeri2008,Drechsler}. Rather, an unconventional pairing
mechanism may be at work in these materials as in the high
temperature superconducting cuprates
\cite{Cruz2008,MonthouxPinesReview}. These compounds are
particularly interesting as they are the first example of
unconventional superconductivity with a large transition
temperature in a non-cuprate transition metal compound. Yet like
the cuprates, there is mounting evidence that the normal state of
these materials can only be described by excitations of strongly
correlated electronic states \cite{HauleKotliarCoherence}.
Furthermore, the superconductivity in these compounds emerges in
close proximity to a magnetic ground state, which suggests that
magnetic correlations may be relevant for superconductivity
\cite{YinPickett,DongSDW,klauss}. These observations highlight the
need for detailed studies of the \LaOFFeAsx\ family in order to
shed important light on the physics of both the cuprates and
strongly correlated superconductors in general
\cite{CurroPuCoGa5}.

In order to investigate the nature of the superconductivity and
the unusual normal state excitations, we have performed \as\
Nuclear Magnetic Resonance (NMR) and Nuclear Quadrupolar Resonance
(NQR) in \LaOFFeAs\ for both random and oriented powder samples.
In the superconducting state, our results reveal the presence of
line nodes in the superconducting density of states.  We find that
the spin lattice relaxation rate, \slrr, varies as $T^3$ for
$0.1T_c < T< T_c$ for $\mathbf{H}_0 \perp c$, which is
characteristic of a d-wave superconductor. We also find that
\slrr\ is anisotropic in the superconducting state, and is
enhanced when $\mathbf{H}_0$ has a component along the $c$-axis.
This result probably reflects the two-dimensional nature of these
materials, which consist of alternating layers of FeAs and
LaO$_{1-x}$F$_x$. In a type-II superconductor, \slrr\ can be
enhanced in applied fields due to the presence of superconducting
vortices \cite{curroNMRY1248,deSotoFluxoids}. Surprisingly we find
no enhancement for $\mathbf{H}_0 \perp c$; this result suggests
that either the vortices do not exist for this orientation, or
they are pinned between the FeAs planes.  In the normal state, we
find that both \slrrt\ and the spin shift, $K_{\rm s}$, are
strongly temperature dependent, and decrease with decreasing
temperature. The temperature dependent Knight shift is similar to
the pseudogap behavior observed in underdoped high temperature
superconductors
\cite{AlloulYBCOspingap,TakigawaONMRinYBCO,Ahilan2008,Nakai2008}.
However, in contrast to the cuprates, \slrrt\ shows no evidence
for antiferromagnetic fluctuations of local 3d spins. In fact, we
find that the Korringa relation is satisfied for the As between
$T_c$ and 300 K, indicating that the dominant relaxation mechanism
is via spin-flip scattering with itinerant quasiparticles, rather
than via transferred coupling to fluctuating Fe 3d moments.

Polycrystalline samples of \LaOFFeAs\ were prepared by standard
methods as described in \cite{zhu2008}. The crystal structure and
the composition were investigated by powder x-ray diffraction and
wavelength-dispersive x-ray spectroscopy (WDX). The magnetic
susceptibility was measured in external fields 10 Oe $<H<$ 70 kOe
using a SQUID magnetometer, and resistance was measured with a
standard 4-point geometry. A value of $T_c \approx$ 26.0 K was
extracted from these measurements. \cite{Luetkens2008} The
material was then ground to a powder with grain size approximately
1-100 $\mu$m, mixed with Stycast 1266 epoxy in a mass ratio of
24:70, and allowed to cure in an external field of 9.2 T. NMR
spectra of the \as\ ($I=3/2$, $\gamma= 7.2917$ MHz/T, 100 \%
abundance) in the random powder were obtained by summing the
Fourier transform of the spin-echo signal as a function of
frequency in a fixed magnetic field of 7.0494 T
\cite{GilClarkFFTSum}.
\begin{figure}
\begin{center}
 \includegraphics[width=70mm,clip]{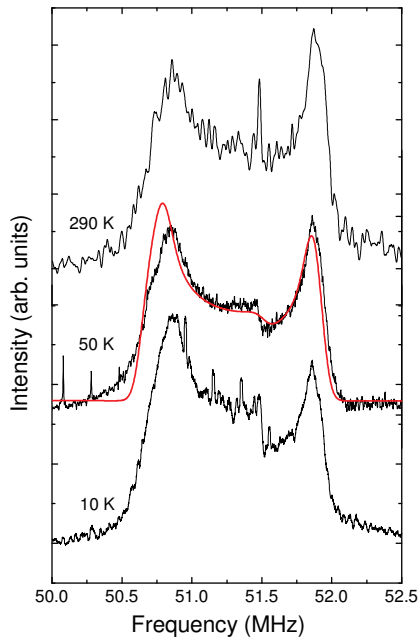}
 \caption{(Color online) Frequency swept \as\ NMR powder spectra of \LaOFFeAs\ measured at  $H_0= 7.0494$ T.
 The solid red line is a simulation of the powder spectrum  as described in the text.}
\label{spectra}
\end{center}
\end{figure}
Typical powder spectra are shown in Fig. \ref{spectra}. \as\ has a
large quadrupolar moment ($Q = 0.3$b) that interacts with the
local electric field gradient (EFG) in the crystal.  The nuclear
spin Hamiltonian is given by: $\mathcal{H} = \gamma\hbar H_0
(1+K)\hat{I}_z + \frac{h\nu_Q}{6}[(3\hat{I}_{z'}^2 - \hat{I}^2) +
\eta(\hat{I}_{x'}^2-\hat{I}_{y'}^2)]$, where $K$ is the magnetic
shift,  $\nu_Q =\frac{3eQV_{zz}}{2I(2I-1)h}\sqrt{1+\eta^2/3}$ is
the NQR frequency, $V_{\alpha\beta}$ is the EFG tensor, and $\eta$
is the asymmetry parameter of the EFG.  Note that the principle
axes $\{ x',y',z' \}$ of the EFG tensor are defined by the local
symmetry in the unit cell, thus the resonance frequency of a
particular nuclear transition, $f(\theta,\phi)$, is a strong
function of field direction relative to the crystalline axes.  For
a powder, the external field is oriented randomly  and the
spectrum is typically quite broad. However, peaks in the powder
pattern correspond to stationary points of the function $f$. Fig.
\ref{spectra} shows the powder pattern central transition
($I_z=1/2 \leftrightarrow -1/2$) of the As, and the two horn peaks
correspond to crystallites with $\theta\approx 41.8^\circ$ (lower
frequency peak), and to $\theta = 90^\circ$ (upper frequency
peak), where $\theta$ is the angle between $z$ and $z'$
\cite{CPSbook}.  The solid red line is a simulation of the powder
pattern including both EFG and anisotropic spin shift effects.  We
find that the spectrum can be fit reasonably well with
$K_a=K_b=0.14$\%, $K_c=0.2$\%, $\nu_Q=10.9$ MHz, and $\eta=0.1$.
We have also done zero-field NQR of the As, and found $\nu_Q =
11.00(5)$ MHz.

\begin{figure}
\begin{center}
 \includegraphics[width=70mm,clip]{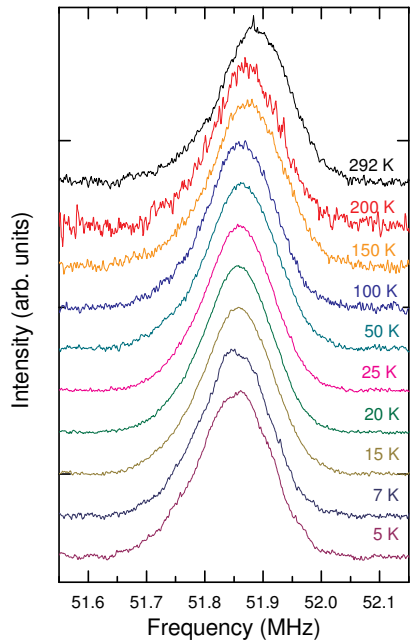}
 \caption{(Color online)  \as\ NMR spectra of the oriented powder measured with $\mathbf{H}_0\perp c$ and $H_0= 7.0494$ T.}
\label{oriented}
\end{center}
\end{figure}

Spectra for the oriented powder are shown in Fig. \ref{oriented}
for $\mathbf{H}_0$ parallel to the orientation direction.
Clearly, the alignment process works well, as the linewidth of the
oriented sample is nearly an order of magnitude smaller and the
powder pattern structure has disappeared. The orientation of the
powder was measured by x-ray diffraction and high field
magnetization measurements. We find that the magnetic
susceptibility is greatest in the plane ($\mathbf{H}_0 || ab$),
where the crystal $a$ and $b$ axes are magnetically equivalent.
The alignment axis of the powder is thus parallel to the $ab$
direction, which is consistent with our observation that the
resonance frequency of the aligned powder corresponds with the
upper horn of the powder pattern. Therefore, the principle axis of
the EFG with the largest eigenvalue must lie along the crystal
$c$-axis. The As has four nearest neighbor Fe atoms, and lies just
above or below the Fe plane. This site is axially symmetric, which
is consistent with our observation that $\eta\approx 0$.

\begin{figure}
\begin{center}
 \includegraphics[width=\columnwidth,clip]{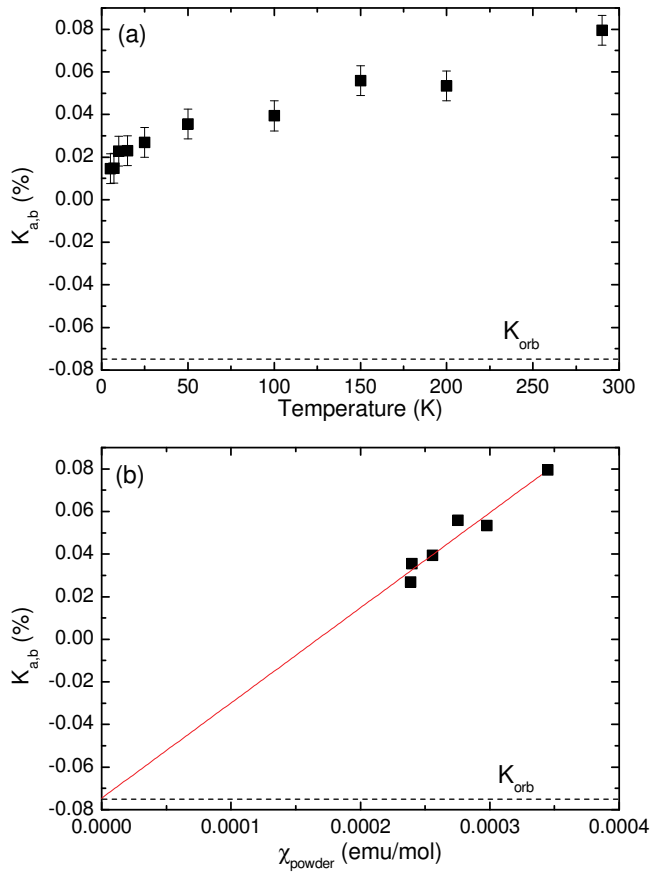}
 \caption{(Color online) (a) Magnetic shift, $K_{ab}$ versus temperature in the aligned sample. The dashed line indicates $K_{\rm orb}$
 (b): $K_{ab}$ versus $\chi_{\rm powder}$, the magnetic susceptibility of
 the powder sample. The solid line is a fit as described in the text.}
 \label{shift}
\end{center}
\end{figure}

The resonance frequency, $f$, of the spectra in Fig.\ref{oriented}
is given by: $f = \gamma H_0(1+K_{ab}) + 3\nu_Q^2/16\gamma H_0$
for $\eta=0$. We have measured that $\nu_Q$ is independent of
temperature, therefore the temperature dependence is entirely due
to the magnetic shift, $K_{ab}$.  Fig. \ref{shift} shows
$K_{ab}(T)$ as a function of temperature, measured along the
orientation axis of the sample.
The magnetic shift arises from the interactions between the As
nuclei and the surrounding electrons; generally it can be divided
into $K_{\rm tot} = K_{\rm s} + K_{\rm orb}$, where $K_{\rm
s}=A\chi_{\rm spin}$ arises via a hyperfine coupling to electron
spins, and $K_{\rm orb}$ arising from orbital magnetization
induced at the nuclear site \cite{CPSbook}.  In order to measure
the hyperfine coupling, we have also measured the bulk magnetic
susceptibility of the powder sample. Fig. \ref{shift}b shows
$K_{ab}$ versus $\chi_{\rm powder}$ with temperature as an
implicit parameter \cite{HajoNote1}. A linear fit to this data
yields a hyperfine coupling $A=25(3)$ kOe/$\mu_B$ plus an offset
of $K_{\rm orb}=-0.075$ \%. Note that this is assuming negligible
diamagnetic and Van Vleck contributions to $\chi_{\rm powder}$. If
these contributions are to represent 50 \% of $\chi_{\rm powder}$,
$K_{\rm orb}$ is reduced to $-0.008$ \% \cite{HajoNote2}. The
striking feature in Fig. \ref{shift}a is a suppression of spin
susceptibility with decreasing temperature in the normal state.
This behavior is identical to that of the pseudogap in the
cuprates, and has been observed in $^{19}$F NMR for $x = 0.11$
\cite{AlloulYBCOspingap,Ahilan2008}. In the superconducting state,
we find that $K_{ab}$ decreases as well, which is suggestive of
spin-singlet pairing.

\begin{figure}
\begin{center}
 \includegraphics[width=\columnwidth,clip]{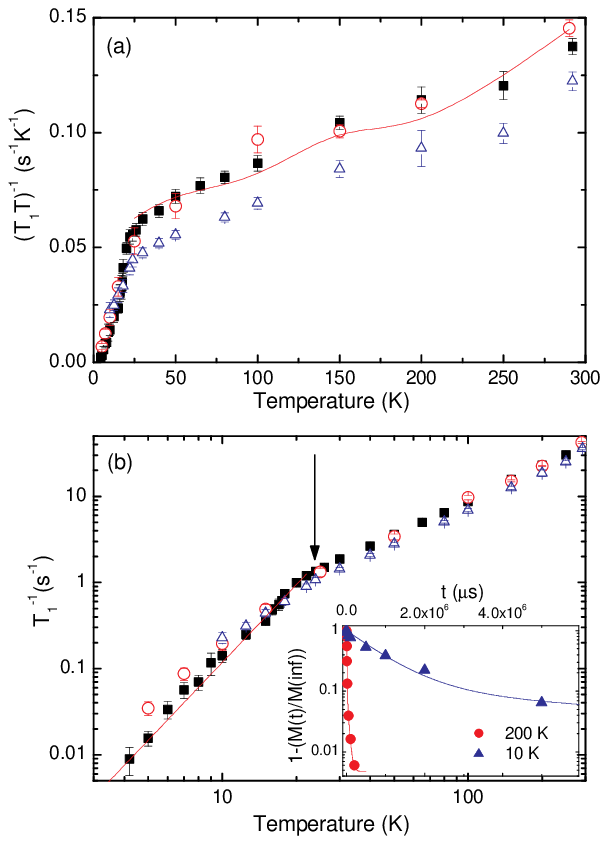}
 \caption{(Color online) (a) The \as\ \slrrt\ versus temperature in \LaOFFeAs\ measured at the upper horn ($\blacksquare$, $\theta=90^\circ$),
 the lower horn ($\triangle$ (blue), $\theta\approx 41.8^\circ$), and in the aligned sample ($\circ$ (red), $\theta=90^\circ$).
 The solid line is given by $K_s^2(T)/\alpha \kappa$ as described in the text.  (b) \slrr\ versus temperature. The arrow indicates $T_c=23.5$ K
 at  7T, and the solid line indicates \slrr$\sim T^3$, indicative of line nodes. Inset: Magnetization recovery and fits for two temperatures.}
\label{T1}
\end{center}
\end{figure}

We have measured the spin-lattice relaxation rate, \slrr, in both
the normal and superconducting states. The relaxation was measured
by inversion recovery of the longitudinal magnetization, and the
data are well fit to the standard expression for the central
transition of a spin $I=3/2$ nucleus with a single $T_1$ component
(see inset of Fig.\ref{T1}). We measured  \slrr\ on both peaks of
the powder sample and on the aligned sample; the data are compared
in Fig. \ref{T1}a. The center frequency of the aligned powder
spectrum and the upper horn of the powder pattern agree well,
supporting our conclusion that the main axis of the EFG is
perpendicular to the alignment axis ($ab$). The \slrrt\ data for
$\theta\approx 41.8^\circ$ agrees reasonably with the data from
Nakai \etal\ \cite{Nakai2008} below $T\approx200$ K. However,
Nakai \etal\ have measured \slrrt\ at the center of the powder
spectrum close to the unshifted frequency. The temperature
dependence of \slrr\ is well described by a Korringa relation
within experimental uncertainty. The solid line in Fig. \ref{T1}a
is a plot of $K_s^2(T)/\alpha\kappa$, where $\kappa$ is the
Korringa constant: $\kappa=\hbar\gamma_e^2/4\pi ~ k_B
^{75}\gamma^2$, $\gamma_e$ is the gyromagnetic ratio of the
electron, $k_B$ is the Boltzmann constant, and $\alpha=1.8$ is a
constant.  A linear scaling (\slrrt $\sim K_s$) was suggested for
$x=0.11$ via $^{19}$F NMR \cite{Ahilan2008}. This behavior is
surprising, since this material is close to an SDW quantum
critical point and fluctuations of Fe 3d moments are a possible
source for spin lattice relaxation \cite{Si2008}. It is not clear
if the hyperfine coupling $A$ represents an on-site contact
interaction, or a transferred interaction to nearest neighbor Fe.
If the coupling is transferred, then the symmetric location of the
As can render it insensitive to magnetic fluctuations at the
antiferromagnetic wavevector; such is the case for the oxygen site
in \ybcox which also shows Korringa behavior even though the Cu 3d
spins are strongly correlated \cite{TakigawaONMRinYBCO}. However,
in the \LaOFFeAsx, Nakai \etal\ have observed a dramatic
enhancement of the As \slrr\ at lower doping levels when the
system undergoes a spin density wave transition \cite{Nakai2008}.
Therefore, the As form factor must not entirely cancel at the
ordering wavevector. In contrast with the cuprates, we find no
evidence of a pseudogap peak in \slrrt\ at a temperature
$T^\star>T_c$ up to 300 K, nor any scaling of \slrrt\ with $T/T_c$
in the normal state
\cite{BarzykinCuprateScaling,CurroPinesFiskMRS}. The pseudogap
peak was used to define the energy scale $T^\star$ of the
pseudogap in underdoped cuprates. It is the temperature where
\slrrt\ of the Cu shows a broad maximum. Our result suggests that
either the As is insensitive to antiferromagnetic fluctuations, or
these fluctuations have disappeared by $x=0.10$.

In the superconducting state, we find that \slrr\ exhibits a drop
at $T_c$, with no evidence for a Hebel-Slichter coherence peak.
For $T\ll T_c$, \slrr\ varies as $T^3$, as seen in Fig. \ref{T1}.
This behavior is indicative of line nodes in the superconducting
gap function, $\Delta(\mathbf{k})$, and contrasts with the
exponential behavior (\slrr $\sim~\exp(-\Delta/k_B T)$) expected
for an isotropic superconducting gap.  In the presence of an
external magnetic field, we find that \slrr\ is anisotropic below
$T_c$.  As seen in Fig. \ref{T1}b, \slrr\ is greater for $\theta
\approx 41.8^\circ$ than for $\theta = 90^\circ$. This result
suggests that in addition to the excited quasiparticles,
superconducting vortices may contribute to \slrr. In particular,
the vortices can give rise to Doppler shifted quasiparticles in
extended states outside the vortex cores, or may be contributing a
dynamical field from simple vortex motion \cite{CurroSlichter}.
The net effect is an increase in \slrr\ above the $T^3$ behavior
expected in zero field. Surprisingly, for $\mathbf{H}_0 \perp c$
we find no such enhancement down to 4 K.  This absence suggests
that the vortices may be pinned between FeAs layers, and therefore
do not contribute to the relaxation rate at the As site. Similar
effects have been observed in other layered superconductors
\cite{deSotoFluxoids}, and may be a natural result of the short
superconducting coherence length, $\xi_c$, along the
$c$-direction.

In summary, we have measured the As NMR and NQR in the normal and
superconducting states of \LaOFFeAs\, and find evidence for line
nodes in the superconducting gap function, and a pseudo-spin gap
in the normal state. This pseudo-spin gap gives rise to a
suppression of $K_s$ and \slrr\ for temperatures below 300 K.

We thank S.-L. Drechsler, W. Pickett and R. Singh for helpful
discussions, and M. Deutschmann, S. M\"uller-Litvanyi, R.
M\"uller, R. Vogel and A. K\"ohler for experimental support. This
work has been supported by the DFG, through FOR 538. G. L.
acknowledges support from the Alexander von Humboldt-Stiftung.


\begin{thebibliography}{24}
\bibitem[{Haj()}]{HajoNote1}
\bibinfo{note}{The oriented sample contained a small paramagnetic impurity
  phase with an increasing susceptibility with decreasing temperature;
  therefore we plot $K_{ab}$ versus $\chi_{\rm powder}$, which contains no
  paramagnetic impurity phase.}
\end{thebibliography}
\end{document}